\newcommand{\Lsun} {L$_{\odot}$}

\documentclass{iaus}
\usepackage{graphicx}

\title[The processing of radiation by dust in galaxies
] 
{The processing of radiation by dust in galaxies}

\author[Ralf Siebenmorgen \& Frank Heymann]   
{Ralf Siebenmorgen$^1$
 \and Frank Heymann$^{1,2}$}

\affiliation{$^1$European Southern Observatory, Karl-Schwarzschild-Str. 2,
        D-85748 Garching b. M\"unchen, Germany
\\ email: {\tt Ralf.Siebenmorgen@eso.org}  
\\[\affilskip]
 $^2$Department of Physics and Astronomy, University of Kentucky,
Lexington, KY 40506-0055, USA 
\\ email: {\tt fheymann@pa.uky.edu}}

\pubyear{2011} 
\volume{284}  
\pagerange{1--12}
\jname{The Spectral Energy Distribution of Galaxies}
\editors{R.J. Tuffs \&  C.C.Popescu, eds.}
\begin{document}

\maketitle

\begin{abstract}
Optical/UV photons and even harder radiation components in galaxies
are absorbed and scattered by dust and re-emitted at infrared
wavelengths. For a better understanding of the obscured regions of the
galaxies detailed models of the interaction of photons with dust
grains and the propagation of light are required. A problem which can
only be solved by means of numerical solution of the radiative
transfer equation. As a prologue we present high angular mid IR
observations of galactic nuclei in the spirit of future ELT
instrumentation. Dust models are discussed, which are suited to fit
the extinction curves and relevant to compute the emission of external
galaxies.  Self-consistent radiative transfer models have been
presented in spherical symmetry for starburst nuclei, in two
dimensions for disk galaxies (spirals) and, more recently, in three
dimensional configuration of the dust density distribution. For the
latter, a highlighting example is the clumpy dust tori around
AGN. Modern advances in the field are reviewed which are either based
on a more detailed physical picture or progress in computational
sciences.

\end{abstract}

\firstsection 
	
\begin{figure}[htb]
\begin{center}
 \includegraphics[width=10.3cm,angle=90]{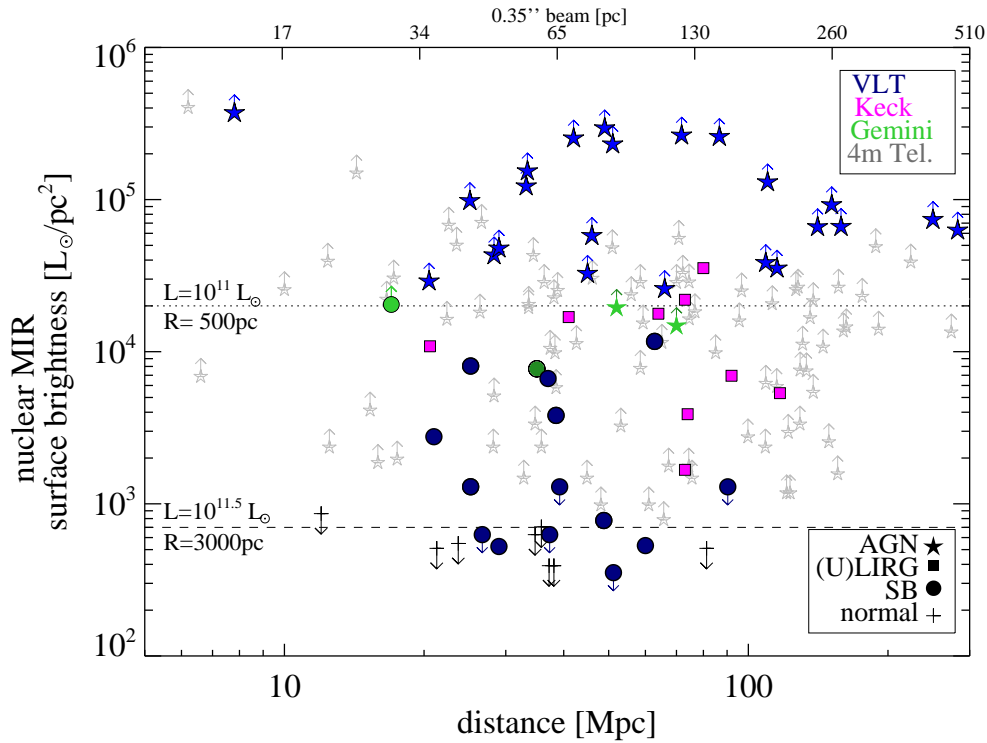} 
\caption{Nuclear mid IR surface brightness versus distance (as plotted
  by Siebenmorgen et al. 2008).  Large symbols mark data from VLT
  (blue): Haas et al. (2007), Horst et al. (2006, 2008), {\itshape
    this work} ; \, Keck (magenta): Soifer et al. (1999, 2000, 2001) ;
  \, Gemini (green): Alonso-Herrero et al. (2006), Mason et
  al. (2007); small symbols from 4\,m class telescopes (gray): Haas et
  al. (2007).  AGN (stars) have $S > 20000$\, \Lsun/pc$^2$ when
  observed with 8\,m class telescopes, starburst and (U)LIRGs are
  below (with the exception of VV114A which may probably harbor an
  AGN). Normal galaxies ($+$) are not detected. The horizontal dotted
  and long-dashed lines give the surface brightness computed with
  starburst models by Siebenmorgen \& Kr\"ugel (2007) for given total
  luminosity $L$ and where stars and dust are distributed in a
  $A_{\rm{V}} = 18$\,mag nucleus of radius $R$.
\label{surf.ps}
}
\end{center}
\end{figure}

\section{Observations}

\noindent
Dust enshrouded activity of a galaxy can be studied ideally by
mid--infrared (MIR) observations. To explore the origin of the nuclear
MIR emission of galaxies as being due to either active galactic nuclei
(AGN) or star formation, observations of high spatial resolution are
required.

\noindent
The nuclear MIR surface brightness is introduced as a quantitative
measurement for AGN and starburst activity. However, one is unable to
distinguish between both activity types using the nuclear MIR surface
brightness derived from 4m class telescopes, even when adopting the
theoretical diffraction limit of $0.7''$ (FWHM) of such telescopes
(cmp. small gray symbols in Fig.~\ref{surf.ps}).  Since the PSF width
is twice as large as for a 8m class telescope and the point source
sensitivity is a factor 16 lower, it becomes more difficult for a 4m
to resolve starburst and the surface brightness of unresolved sources
is reduced. Data recently obtained at 8m class telescopes
(Siebenmorgen et al. 2008) show that, out to a distance of 100Mpc, the
MIR surface brightness acquired clearly differentiate AGN from SB
behavior (Fig.~\ref{surf.ps}). Utilizing VISIR at the VLT the AGN
still appear point like whereas most starburst are resolved in the
MIR.  This discrimination was made possible by an increase in spatial
resolution by a factor 2. Therefore it provides a clue to what will be
possible by increasing the spatial resolution by another factor 5 when
going from the VLT to the proposed extreme large telescope such as the
E-ELT which will be 40m class. For the E-ELT a mid infrared instrument
is included in the instrumentation plan and it has beside imaging also
high resolution spectroscopic and polarimetric observing capabilities
(Brandl et al., 2010).

\begin{figure}[htb]
\begin{center}
\includegraphics[width=11cm,angle=0]{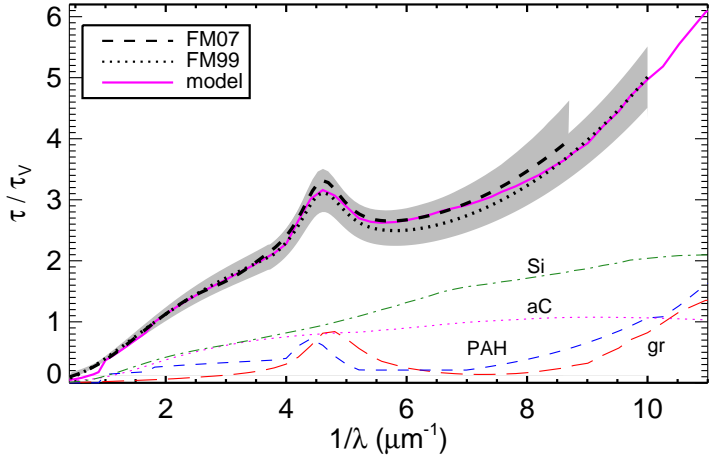}
\caption{Mean extinction curve of the ISM by Fritzpatrick\&Massa
  (2007, black) and a fit (magenta) by the dust model of
  Sect.~\ref{dust}. Individual dust components are shown as
  labeled. The grey areas indicate the 1$\sigma$ deviation as of the
  samples.
\label{dust.ps} }
\end{center}
\end{figure}

\section{Dust model \label{dust}}

\noindent 
Teams interested in modelling the processing of radiation by dust in
galaxies often apply a dust model as derived for the diffuse ISM of
the Milky Way.  Dust cross sections are computed using similar optical
constants and temperature fluctuating particles such as PAHs are
included.  We developed one of such dust models in which {\it {large}}
($60\rm{\AA}<a<0.2-0.3\mu$m) silicate (Draine 2003) and amorphous
carbon (Zubko et al. 2004) grains and {\it {small}} graphite
($5\rm{\AA}<a<80$\,\AA) grains are considered.  We apply a power law
size distribution: $n(a) \propto a^{-3.5}$ and absorption and
scattering cross-sections are computed with Mie theory.  In addition
there are {\it {PAHs}} with 30 and 200 C atoms with absorption cross
section as given by Schutte et al. (1993).  By computing cross
sections above 100\,eV, we consider an approximation of kinetic energy
losses (Dwek\& Smith 1996) and apply it to all particles. The choice
of parameters is set up to achieve a fit of the mean extinction curve
of the ISM (Fitzpatrick \& Massa 2007), as shown in Fig.\ref{dust.ps}.

\noindent 
With the advent of {\it{ISO}} and {\it{Spitzer}} more PAH emission
features and more details of their band structures are detected
(Tielens 2008).  We consider 17 emission bands and take Lorentzian
profiles (Siebenmorgen et al. 1998). Parameters of what we call
astronomical PAH are calibrated using mid-IR spectra of starburst
nuclei and the RT model as of Sect.~{\ref{sb}}. PAH cross sections of
the emission bands are listed by Siebenmorgen \& Kr\"ugel (2001). In
the model, we use dust abundances of [X]/[H] (ppm) of: 31 for [Si],
150 [amorphous C], 50 [graphite] and 30 [PAH], respectively; which is
in agreement with cosmic abundance constraints (Asplund et
al. 2009). We are in the process of upgrading the model to be
consistent with the polarization of the ISM (Voshchinnikov
2004). Besides extinction the model accounts for the diffuse emission
of solar neighborhood when the dust is heated by the interstellar
radiation field (Mathis et al. 1983).

\section{Starbursts \label{sb}}

\noindent 
There are three different ways in the literature trying to reproduce
the SED of extra-galactic nuclei.  A first one uses a SED of a
well-known galaxy as a template to match other objects (Lutz et
al. 2002). A second group reproduce the shape of the SED by optical
thin dust emission using a scaled up version of the interstellar
radiation field (Draine \& Lee 2007). A third and more ambitious
method is to solve the radiative transfer (RT) problem using
assumptions about the galaxy.  The latter is done at various levels of
sophistication and it may be instructive to point out technical
differences. Teams solving the RT problem evaluate the emission from a
dusty medium of spheroidal shape filled with stars and dust.  At first
glance, the model results appear to agree, but upon closer inspection
one finds that deviations of derived parameters are substantial.  For
example for Arp220 Grooves et al.  finds an optical depth of a few
whereas we derive values between 70 -- 120.  We admit that we did not
always find it easy to pin down exactly which approximations our
colleagues used, still we try to summarize the main features of some
RT models which are in widespread use.  Monte Carlo techniques are
discussed in Sect.~\ref{agn}.  We use the term {\it {dust
    self--absorption}} when photons which are emitted by a dust
particles may be absorbed by other dust particles within the model
sphere and we call {\it {exact RT}} when the RT equation is solved
accurately including multiple scattering and dust self--absorption.

{$-$} Groves (2004, this volume): Shock, photoionization and dust
radiative transfer code called MAPPINGIII. The dust is distributed in
a screen and the RT is solved in one dimension ignoring dust
self-absorption and treating scattering in forward direction only.

{$-$} Efstathiou \& Rowan--Robinson (2003), Efstathiou (this volume):
Dust is distributed in spherical symmetry. There are two components:
molecular clouds with exact RT computation and cirrus which is added
as a foreground screen. Both components are uncoupled in the RT
equation. The code includes a free parameter to scope with the
observed optical and UV spectrum.

{$-$} Takagi et al. (2003, this volume): Exact RT in spherical
symmetry where dust and stars are homogeneously distributed. Molecular
clouds (clumps) are not treated. The code includes a free parameter to
scope with the observed optical and UV spectrum.

{$-$} Siebenmorgen et al. (2007): Exact RT in spherical symmetry where
dust and a young and old population of stars are distributed in the
galaxy. A fraction of the young stars are in molecular clouds for
which a second exact RT computation is solved. The coupling of both RT
solutions, that of the galaxy and the embedded sources, is treated and
this is a particular feature of the model.

{$-$} Silva et al. (1998): Present a code called GRASIL in which dust
is distributed in axial-symmetry.  There is a molecular cloud
component with exact RT. A cirrus component is added in which the RT
is solved by ignoring dust self-absorption and in which dust
scattering is simplified by altering the dust absorption cross
section. In addition the code includes a free parameter to scope with
the observed optical and UV spectrum. All three components are
uncoupled in the RT equation.

{$-$} Popescu et al. (2002, this volume): Dust is distributed in
axial-symmetry.  The model is fine tuned to Spiral galaxies
(Sect.~\ref{spirals}) and includes a bulge, two galactic disk
components and molecular clouds (clumps).  The RT is solved by
ignoring dust self-absorption and only the first scattering event is
treated. Clumps are added by a pre-computed template spectrum.  There
is a free parameter to scope with the observed optical and UV
spectrum. All five components are uncoupled in the RT equation.

\begin{figure}[htb]
\begin{center}
\includegraphics[width=8cm,angle=90]{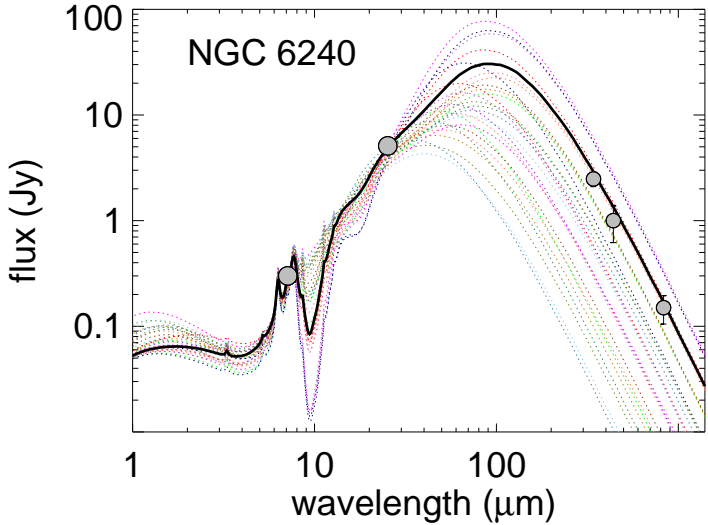}
\caption{Elements of the SED starburst library by Siebenmorgen et
  al. (2007, dotted) which fit photometry (circles) of NGC6240 at
  8$\mu$m (Siebenmorgen et al. 2004) and 24$\mu$m (Klaas et al. 2001)
  to within 30\%; submm data by Benford (1999) and Klaas et
  al. (2001), best fit is indicated as thick line \label{2dat.ps}. }
\end{center}
\end{figure}

\noindent 
As the coupling between the RT calculations of the galaxy and the
embedded sources is a computer intensive problem we provide a library
of some 7000 SEDs {\footnote{SED library available at: {\tt
      {http://www.eso.org/$\sim$rsiebenm/sb\_models/}}}} for the
nuclei of starburst and ultra--luminous galaxies (Siebenmorgen et al.,
2007).  Its purpose is to quickly obtain estimates of the basic
parameters of the object, such as luminosity, size and dust or gas
mass and to predict the flux at yet unobserved wavelengths using a
physical model. Unfortunately, for faint or red-shifted objects
photometry is sometimes only provided at two MIR bands, for example at
8 and 24$\mu$m from Spitzer. In Fig.~\ref{2dat.ps} we demonstrate for
the ULIRG NGC6240 that the SED library can be used to estimate the
total luminosity to within a factor $\sim 2$ in case only two such MIR
fluxes are known.  We also see (Fig.~\ref{2dat.ps}) that the SED will
be quite well constrained by an additional submm data point.  The
model is applied to high red shifts ($z \approx 3$, Efstathiou \&
Siebenmorgen 2009) where PAH have been detected.

\section{Spirals \label{spirals}}

\begin{figure}[htb]
\begin{center}
\includegraphics[width=10.5cm,angle=270]{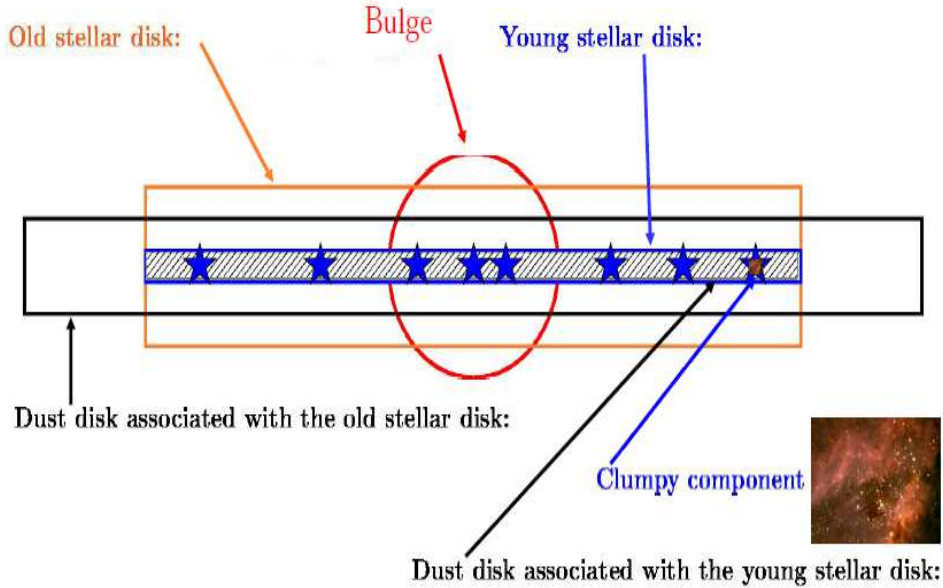}
\end{center}
\vspace*{-1.0 cm}
\caption{Schematic view of the geometry applied in the models
of Spirals by Popescu et al. \label{scheSpiral.ps} }

\end{figure}

\noindent
The geometrical distribution of dust in spiral galaxies has been
investigated by fitting images in the optical/NIR.  This is done by
means of 2d radiative transfer codes which include a bulge component
and an old stellar population of stars; originally only dust
absorption and scattering was treated (Kylafis \& Bahcall 1987,
Xilouris et al. 1999, Misiriotis et al., 2000). Building into the
models typical dust emission properties gives a puzzle when compared
to observations in the FIR. Typically the models underestimate the FIR
luminosities by a factor of 3.  The FIR excess could be explained by
Popescu et al. (2000, 2010, this volume) introducing more components.
In their models there is a distribution of diffuse dust associated
with the old and young stellar disk populations as well as a clumpy
component arising from dust in the parent molecular clouds in star
forming regions. Basic parameters of their models are: angular size
and inclination of the disk, the central face-on dust opacity in the
B-band, a clumpiness factor for the star-forming regions, the
star-formation rate, the normalized luminosity of the stellar
population and the bulge-to-disk ratio and a wavelength dependent
escape probability of stellar radiation (which is sometimes treated as
model output). A schematic view of the geometry of the RT model is
presented in Fig.~\ref{scheSpiral.ps}. The model is successfully
applied to a large sample of galaxies from the Millennium Galaxy
Catalog Survey (Driver et al. 2007). The observed
attenuation--inclination relation is fit when a two dust disks are
considered whereas data are not fit in a single disk scenario.

Over the past decade the edge on spiral NGC891 become a benchmark test
for RT models of spiral galaxies. Bianchi (2008) uses a clumpy disks
model and applied MC techniques which are of advantage when dealing
with such complicated geometries (see Sect.~\ref{agn}). His results
give support to the idea that the diffuse dust disk is more extended
than the stellar disks. De Looze et al. (20011, this volume) present a
MC radiative transfer model of the Sombrero galaxy which is able to
fit the SED and optical/NIR images and extinction profiles. Using only
an old stellar population the dust luminosity in the FIR is again
underestimated by a factor 3.  The discrepancy is solved by assuming a
star forming stellar component both in the ring and inner disk to
account for emission at 24 and 70$\mu$m. In the submm an additional
dust component is used, accounting for 75\% of the total dust content
and distributed in quiescent compact clumps. A possibility that part
of the dust could be composed of grains with a higher submm
emissivity, of for example large fluffy grains (Kr\"ugel \&
Siebenmorgen 1994), could not be ruled out. On the other hand a
similar approach introducing a clumpy medium in a single disk model is
applied to the edge-on spiral galaxy UGC4754 (Baes et al. 2011). The
model has a deficit in the FIR luminosity and the authors propose
higher submm dust emissivities to solve the energy ballance at such
long wavelengths.

\section{AGN and Monte Carlo \label{agn}}

\noindent
Dust is detected in the majority of active galactic nuclei (Haas et
al., 2008).  According to the unified scheme (Antonucci \& Miller,
1985), AGN are surrounded by a dust obscuring torus. However,
observations are not able to resolve the inner parts of AGNs so that
the geometrical distribution of the dust is a matter of
debate. Theoretical considerations favor a clumpy structure in a torus
like configuration of optical thick dust clouds surrounding the black
hole accretion disk (Pier \& Krolik (1992). Some evidence of a clumpy
or filamentary structure of the torus is given by VLTI observations of
the nearby active galactic nuclei in Circinus (Tristram et al. 2007).
Radiative transfer models of homogeneous toroidal structures
over-predict the silicate absorption and emission band at around
10$\mu$m when compared to observations. The silicate emission feature
in type I AGN is rather shallow (Siebenmorgen et al. 2005). A
statistical attempt to describe the radiation from a clumpy AGN torus
is given by Nenkova et al. (2002) and more detailed radiative transfer
computations using the Monte Carlo technique are presented by H\"onig
et al. (2006) and Schartmann et al. (2008).

We develop a vectorized three dimensional Monte Carlo (MC) technique
to solve the radiative transfer problem in such a fairly complicated
geometry of a clumpy dust torus. Originally the MC radiative transfer
method for scattered light in astronomical sources is presented by
Witt (1977) and important improvements are introduced by Lucy
(1999). A first version of the code which we are using is developed by
Kr\"ugel (2006). To reduce the computational effort of MC dust
radiative transfer methods different optimization strategies are
developed (Lucy 1999, Bjorkman \& Wood 2001, Gordon et al. 2001,
Misselt et al. 2001, Baes 2008, Bianchi 2008, Baes et al. 2011). A
numerical solution of the radiative transfer equation which is
specifically developed to be vectorized and to run on graphical
computer units (GPU) is presented by Heymann (2010). A GPU version
including stochastic heated grains is presented (Siebenmorgen et
al. 2010), in which some detailed physics of the destruction of PAHs
by soft (photospheric) and hard (X--ray) radiation components is
treated, (see Siebenmorgen \& Kr\"ugel 2010 for a discussion of PAH
dissociation).  The speed up factor of the GPU method is proportional
to the number of processors available. Therefore the GPU scheme is on
our standard PC with a conventional graphic card about 100 times
faster when compared to the original scalar version of the MC program.
The numerical solution provides the temperature of the dust, spectra
and images within acceptable timescales. The code is tested against
existing benchmark results. The method handles arbitrary dust
distributions in a three dimensional Cartesian model space at various
optical depths. Therefore it is well adopted to be applied to a clumpy
torus structure around an AGN.

\begin{figure}[htb]
\begin{center}
\includegraphics[width=6cm,angle=0]{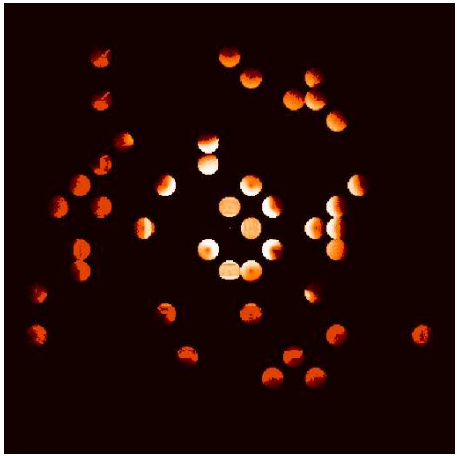}
\end{center}
\caption{Zoom into the clumpy AGN torus displaying the temperature
  structure of individual clouds. \label{agnTklump} }
\end{figure}

\begin{figure}[htb]
\begin{center}
 \includegraphics[width=13.5cm,angle=0]{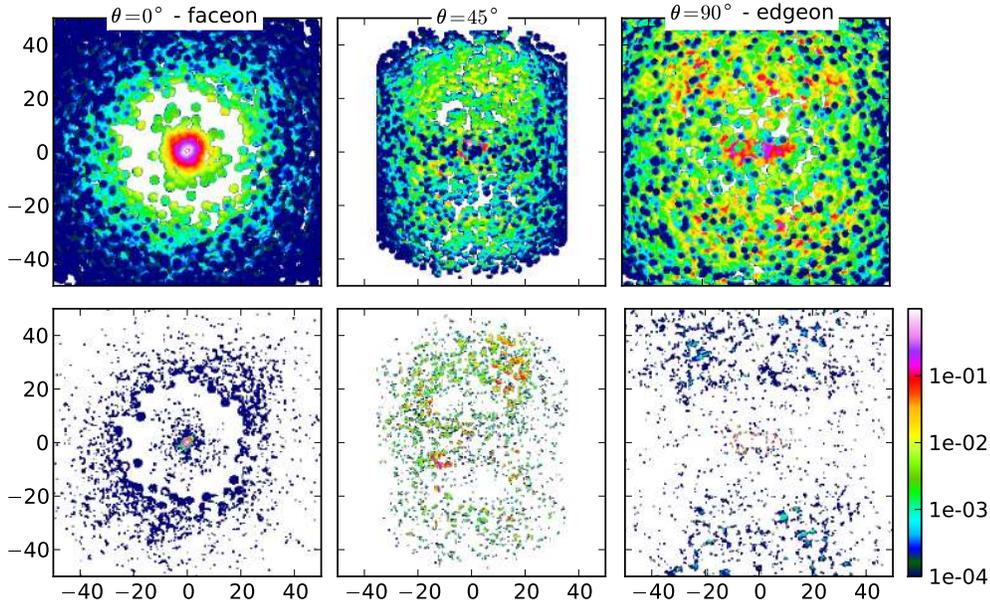}
\caption{Intensity maps of the clumpy AGN dust torus model
  (color-code: cgs units normalized to the peak intensity).  The model
  is made up of 1000 clumps and each clump has a total optical depth
  of $\tau_{\rm V} = 30$.  Mid IR images at 10\,$\mu$m are dominated
  by dust emission and displayed in the top panel. The optical light
  at 0.55\,$\mu$m represents the scattering intensities and are shown
  in the bottom panel.  The AGN torus is viewed face-on ($\theta =
  0^\circ$, left), at $\theta = 45^\circ$ (middle) and edge-on
  ($\theta = 90^\circ$, right).  \label{agn.ps} } 
\end{center}
\end{figure}

The primary heating source of the dust in the torus emerges from the
accretion disc around the massive black hole. For the spectral shape
we apply a broken power law as suggested by Rowan-Robinson (1995). As
example we use a total AGN luminosity of $L_{\rm {AGN}} =
10^{45}$\,erg/s.  This sets the inner radius $R_{\rm {in}} \sim 0.4
\sqrt{L_{\rm {AGN}}/10^{45}}$\,[pc] where the dust evaporation zone is
located. We treat the torus to an outer radius of $R_{\rm {out}} = 50
R_{\rm {in}}$ and consider a torus opening angle of $θ \sim 20^{\rm
  o}$.  Clouds are randomly distributed in the otherwise optical thin
dust torus. Individual clouds are approximated with a sphere of
constant density and radius of 3\,pc. The optical depth through the
center of the clouds is $\tau_{\rm V} = 30$. In Fig.~{\ref{agnTklump}}
we zoom into a part of the clumpy torus and show the temperature
structure of the clouds. The individual hemispheres which are oriented
in direction of the central heating source are warmer as compared to
their dark sides. Shadowed clumps become also colder than those
directly heated by the central engine. Intensity maps of the emission
at 10$\mu$m and scattered light at 0.5$\mu$m of the clumpy AGN dust
torus model are displayed in Fig.~\ref{agn.ps}. The maps show that in
a clumpy medium even in the edge on case some radiation from the
central region is visible.  A more detailed description of the clumpy
AGN torus model and a further application to the silicate emission in
Quasars is given in Heymann \& Siebenmorgen (2011).

\end{document}